\documentclass[aps,pre,twocolumn,floatfix,longbibliography]{revtex4-1}
\pdfoutput=1
\usepackage{amsmath}
\usepackage{amssymb}
\usepackage{amsfonts}
\usepackage{graphicx}
\usepackage{bbold}
\usepackage{bm}
\usepackage{bm}
\usepackage{times,float}
\usepackage{graphicx}
\usepackage[usenames,dvipsnames,svgnames]{xcolor}
\usepackage{hyperref}
\hypersetup{colorlinks=true, linkcolor=NavyBlue, citecolor=PineGreen,urlcolor=cyan}
\usepackage[utf8]{inputenc}

\definecolor{red}{rgb}{1.0,0.0,0.0}
\definecolor{blue}{rgb}{0.0,0.0,1}

\newcommand{\bea}{\begin{eqnarray}}
\newcommand{\eea}{\end{eqnarray}}
\newcommand{\nn}{\nonumber}
\newcommand{\etal}{\textit{et al.}}

\usepackage{soul}
\hypersetup{colorlinks=true, linkcolor=NavyBlue, citecolor=PineGreen,urlcolor=cyan}
\DeclareUnicodeCharacter{2212}{-}
\begin{document}

\title{Comments on {\it Superstatistical properties of the one-dimensional Dirac oscillator} by Abdelmalek Boumali \etal}

\author{Jorge David Casta\~no-Yepes$^1$, I. A. Lujan-Cabrera$^{2}$ and C. F. Ramirez-Gutierrez$^{3}$}
\address{
  $^1$Instituto de Ciencias
  Nucleares, Universidad Nacional Aut\'onoma de M\'exico, Apartado
  Postal 70-543, M\'exico Distrito Federal 04510,
  Mexico.\\
  $^2$Ingenier\'ia F\'isica, Facultad de Ingenier\'ia, Universidad Aut\'onoma de Quer\'etaro, C.P. 76010 Quer\'etaro, Qro., Mexico.\\
  $^3$Universidad Polit\'ecnica de Quer\'etaro, El Marqu\'es, Qro., M\'exico}
%
\begin{abstract}
In this comment, we discuss the mathematical formalism used in  Boumali \etal (2020) which describes the superstatistical thermal properties of a one-dimensional Dirac oscillator. In particular, we point out the importance of maintaining the Legendre structure unaltered to ensure an accurate description of the thermodynamic observables when a Tsallis-like statistical description is assumed. Also, we remark that all the negative poles have to take into account to calculate the Gibbs--Boltzmann partition function. Our findings show that the divergences obtained by the authors in the Helmholtz free energy, which are propagated to the other thermal properties, are a consequence of an incomplete partition function. Moreover, we prove that the restrictions over the $q$-parameter are no needed if an appropriate partition function describes the system.
\end{abstract}
\maketitle
\section{Introduction}

 The recently published paper by A. Boumali
\etal~\cite{boumali2020superstatistical} shows the calculation of the thermal
properties of a one-dimensional Dirac oscillator, in the framework of the
superstatistics theory using the Gamma function as the distribution for the
fluctuating inverse temperature. They computed the Helmholtz free energy,
average energy, entropy, and specific heat capacity from an extension of the
canonical formalism, i.e.,~by performing derivatives on the natural logarithm
of the partition function $\mathcal{Z}$. By applying the well-known Beck and Cohen
expansion, the authors write the superstatistical partition function $\mathcal{Z}_{a}$
in terms of the Boltzmann--Gibbs partition function $\mathcal{Z}_0$. The latter is
computed by an analytical treatment of the Cahen-Mellin integral
transformation. In this comment, we revised their calculations, which in our
opinion, do not have a rigorous treatment. We list the main aspects to improve:
\begin{enumerate}
    \item The authors assume constraints on the super statistical distribution function's free parameters, which leads to a Tsallis-like modified Boltzmann factor. By demanding a Legendre structure for that particular kind of statistics, we demonstrate that the divergences on the original manuscript's thermodynamic functions are eliminated. Such a Legendre structure is achieved by considering the $q$-logarithm instead of the natural logarithm.
    
    \item The expansion of the partition function in terms of powers of the parameter $q$ is incomplete.
    
    \item The canonical partition function $\mathcal{Z}_0$ presented by the authors is incomplete, given that in the Cahen-Mellin integral transformation, they have ignored all the poles located at the negative real axis. 
\end{enumerate}

The corrections above impact directly in the thermodynamic functions. In particular, we center our discussion into the specific heat, which is not positive definite in the author's description.

 \section{The superstatistical formalism revisited}
\label{Sec:The_superstatistical_formalism_revisited}

The  work published by Casta\~no \etal~\cite{castano2019super}, emphasizes the
importance of maintaining the Legendre structure of the potentials for a super
statistical description related to a Tsallis non-extensive framework, i.e.,~the
thermodynamic functions can be calculated from the derivatives of the logarithm
of the partition function. In particular, when the Gamma distribution function
models the fluctuation in the intensive parameter $\bar{\beta}$, the Tsallis
formalism for non-extensive thermodynamics appears, so that the
superstatistical partition function reads:
\bea
Z_q(\beta)\equiv \sum_n\left[1-(1-q)\beta E_n\right]^{1/(1-q)},
\label{Zq}
\eea
where $q$ is a parameter which takes into account the grade of
non-extensivity~\cite{tsallis1998role,tsallis1994numbers}, and to simplify the
notation we set $\beta=\langle\beta\rangle$. The Legendre structure is given in terms of the
$q$-logarithm
\bea
\ln_qx\equiv\frac{x^{1-q}-1}{1-q},
\label{lnq}
\eea
which meas that the thermodynamic functions can be written as:
\begin{subequations}
\bea
F_q(\beta)\equiv U_q(\beta)-TS_q=-\frac{1}{\beta}\ln_qZ_q(\beta),
\eea
\bea
U_q(\beta)=-\frac{\partial}{\partial\beta}\ln_qZ_q(\beta),
\eea
\bea
C_q(\beta)=\frac{\partial U_q(\beta)}{\partial T},
\label{CvZ2}
\eea
and
\bea
S_q=k_B\frac{1}{q-1}\left(1-\sum_n p_n^q\right)\forall\;q \in\mathbb{R}.
\label{TsallisEntropy}
\eea
\label{LegendreStructure}
\end{subequations}

As is shown in Ref.~\cite{castano2019super},  the use of $\ln\mathcal{Z}$ instead of
$\ln_q\mathcal{Z}$ in the presented Legendre structure implies that the specific heat
becomes negative, and spurious phase transitions may occur.\\

It is worth mentioning that, in general, the $\chi^2$-superstatistics is not always related to the Tsallis non-extensive formalism. Formally, the distribution function is given by
\bea
   f(\beta)=\frac{1}{b\Gamma(c)}\left(\frac{\beta}{b}\right)^{c-1}e^{-\beta/b},
\eea
from which the Tsallis partition function is found by demanding that the random
variable $\beta$ is positive definite, and the free parameters take the
form $c=1/(q-1)$, and $bc=\beta_0$~\cite{beck2003superstatistics}. Clearly, the
parameters are arbitrary, but the authors select the named constrains, and that
is the reason to guide our comment into a Tsallis-statistics framework. With
that in mind, as it was discussed in Ref.~\cite{wada2002thermodynamic}, the
stability criteria for the Tsallis statistics are the entropy concavity as a
function of the internal energy and the specific heat positivity. Note that the
stability discussion is given in terms of the Tsallis-``corrected'' energy
constrain~\cite{tsallis1998role}:
\bea
   U_q=\sum_n p_n^q E_n,
   \label{UTsallisConstrain}
\eea
where
\bea
   p_n=\frac{\left[1-(q-1)\beta E_n\right]^{1/(q-1)}}{\mathcal{Z}_q}, 
   \label{Tsallispn}
\eea
which have to be satisfied in order to applying the $\chi^2$-Boltzmann factor with the parameters $b$ and $c$ taken with the discussed form. In that sense, the authors' thermodynamic (Legendre) structure does not have a closed-form.  That is easily proved by computing the mean energy prescription:
\bea
   U_q&=&-\frac{\partial}{\partial\beta}\ln\sum_n\left[1-(q-1)\beta E_n\right]^{1/(q-1)}\nn\\
   &=&\frac{\sum_n \left[1-(q-1)\beta E_n\right]^{1/(q-1)-1}E_n}{\sum_n \left[1-(q-1)\beta E_n\right]^{1/(q-1)}},
   \label{prescriptionBoumali}
   \eea
which if the Tsallis probability function of Eq.~\ref{Tsallispn} is used yields
\bea
   U_q=\sum_n p_n E_n\left[1-(q-1)\beta E_n\right]^{-1},
\eea
or by analogy with Eq.~\ref{UTsallisConstrain} gets:
\bea
   U_q=\mathcal{Z}^{q-1}\sum_n p_n^q E_n.
\eea

Therefore, the prescription of Eq.~\ref{prescriptionBoumali} neither resembles the constraint of Eq.~\ref{UTsallisConstrain} nor the maximization condition given by
\bea
   U_q=\sum_n p_n E_n,
\eea
in such a way that the ``average'' interpretation is missed or needs to be clarified. A straightforward calculation shows that the $\ln_q x$ conserves the Legendre structure and the mean energy constrain of Eq.~\ref{UTsallisConstrain}.

\subsection{Expansion of the partition function}

By defining the parameter $a=q-1$, the authors used Beck--Cohen expansion
(around $a=0$)~\cite{beck2003superstatistics}:
\bea
\mathcal{Z}_a\approx\sum_{n}\left[1+\frac{a}{2}\beta^2E_n^2-\frac{a^2}{3}\beta^3E_n^3\right]e^{-\beta E_n},
\eea
which from the fact that
\bea
\beta^k E_n^k=(-1)^k\frac{\beta^k}{\mathcal{Z}_0}\frac{\partial^k\mathcal{Z}_0}{\partial\beta^k},\;\;\mathcal{Z}_0=\sum_n e^{-\beta E_n},
\eea
can be written as
\bea
\mathcal{Z}_a\approx\left[1+\frac{a}{2}\frac{\beta^2}{\mathcal{Z}_0}\frac{\partial^2 \mathcal{Z}_0}{\partial \beta^2}+\frac{a^2}{3}\frac{\beta^3}{\mathcal{Z}_0}\frac{\partial^3 \mathcal{Z}_0}{\partial \beta^3}\right]\mathcal{Z}_0,
\eea
which its third term has a different sign compared with Eq.~(26) of the named
paper. Moreover, the formula presented in Ref.~\cite{beck2003superstatistics}
is not complete in the sense that in such work, Beck and Cohen want to discuss
the equality of the first expansion term for all the distribution functions.
Thus, a complete expansion at order $\mathcal{O}(a^2)$ takes the form:
\bea
\mathcal{Z}_a&\approx&\left[1+\frac{a}{2}\frac{\beta^2}{\mathcal{Z}_0}\frac{\partial^2 \mathcal{Z}_0}{\partial \beta^2}\right.\nn\\
&+&\left.a^2\left(\frac{1}{3}\frac{\beta^3}{\mathcal{Z}_0}\frac{\partial^3 \mathcal{Z}_0}{\partial \beta^3}+\frac{1}{8}\frac{\beta^4}{\mathcal{Z}_0}\frac{\partial^4 \mathcal{Z}_0}{\partial \beta^4}\right)\right]\mathcal{Z}_0,
\eea
which constitutes our first correction.

Our second correction is about the logarithmic prescription. The authors calculate the thermal functions from derivatives of
\bea
\ln\mathcal{Z}&=&\ln\mathcal{Z}_0+\ln\left[1+\frac{a}{2}\frac{\beta^2}{\mathcal{Z}_0}\frac{\partial^2 \mathcal{Z}_0}{\partial \beta^2}\right.\nn\\
&+&\left.a^2\left(\frac{1}{3}\frac{\beta^3}{\mathcal{Z}_0}\frac{\partial^3 \mathcal{Z}_0}{\partial \beta^3}+\frac{1}{8}\frac{\beta^4}{\mathcal{Z}_0}\frac{\partial^4 \mathcal{Z}_0}{\partial \beta^4}\right)\right]\nn\\
&\approx&\ln\mathcal{Z}_0+\frac{a}{2}\frac{\beta^2}{\mathcal{Z}_0}\frac{\partial^2 \mathcal{Z}_0}{\partial \beta^2}+\frac{a^2}{3}\frac{\beta^3}{\mathcal{Z}_0}\frac{\partial^3 \mathcal{Z}_0}{\partial \beta^3}.
\eea

Instead of that, we propose the expansion of $\ln_a\mathcal{Z}$ around $a=0$ given by
\bea
\ln_a Z&=&\ln\mathcal{Z}_0 +\frac{1}{2}a\left(\frac{\beta^2}{\mathcal{Z}_0}\frac{\partial^2 \mathcal{Z}_0}{\partial \beta^2}-\ln^2\mathcal{Z}_0\right)\\
&+&\frac{1}{6}a^2\left(2\frac{\beta^3}{\mathcal{Z}_0}\frac{\partial^3 \mathcal{Z}_0}{\partial \beta^3}-3\ln \mathcal{Z}_0 \frac{\beta^2}{\mathcal{Z}_0}\frac{\partial^2 \mathcal{Z}_0}{\partial \beta^2}+\ln^3 \mathcal{Z}_0\right)\nn.
\label{LnqZ_expansion}
\eea

For particular interest, the specific heat can be written as
\bea
C_v&=&C_v^0+\frac{a}{2}\beta^2\frac{\partial^2}{\partial \beta^2}\Big[C_v^0+k_B\left(\beta U_0\right)^2-k_B\left(\beta F_0\right)^2\Big]\nn\\
&+&\frac{a}{6}\beta^2\frac{\partial^2}{\partial \beta^2}\Bigg[3\beta F_0\left(C_v^0+k_B\left(\beta U_0\right)^2\right)\\
&-&2k_B\beta^3\left(U_0^3+\frac{\partial^2U_0}{\partial\beta^2}\right)-6\beta U_0 C_v^0-k_B\left(\beta F_0\right)^3\Bigg]\nn.
\label{CvSSexpanded}
\eea

Note that the above expression has an involved structure with terms that cannot be easily ignored. Such terms play a crucial role in the desired behavior for the thermal functions, such as the positive definite specific heat.

\section{Calculation of $\mathcal{Z}_0$}
The authors calculate the partition function $\mathcal{Z}_0$ starting from the positive energy spectrum of the 1D-Dirac oscillator~\cite{szmytkowski2001completeness}:
\bea
E_n=mc^2\sqrt{1+2rn},\;\;n=0,1,2, \ldots,
\label{energylevels}
\eea
where $r=\hbar\omega/mc^2$. To perform the summation, they call the well-known Cahen-Mellin transformation in the form:
\bea
e^{-x}=\frac{1}{2\pi i}\int_C\,ds\,\Gamma(s)\,x^{-s},
\eea
so that, by identifying $x=\sqrt{2r}\beta mc^2$, is straightforward to get
\bea
\mathcal{Z}_0&=&\sum_{n=0} e^{-\beta E_n}\\
&=&\frac{1}{2\pi i}\int_C\,ds\,\left(\sqrt{2r}\beta mc^2\right)^{-s}\Gamma(s)\,\zeta_H\left(\frac{s}{2},\frac{1}{2r}\right),\nn
\label{intens}
\eea
where $\zeta_H(s,v)$ is the Hurwitz-zeta function. To perform the integral, the
authors use the Cauchy's residue theorem by identifying such residues at
$s=0$ and $s=2$. However, as it is established in
Ref.~\cite{frassino2020thermodynamics}, the integration limits have the form:
\bea
\int_C\rightarrow\int_{\mathfrak{c}-i\infty}^{\mathfrak{c}+i\infty}\;\;\text{with}\;\;\mathfrak{c}\in\mathbb{R}^+.
\eea
\begin{figure}[h]
  \begin{center}
    \includegraphics[width=0.3\textwidth]{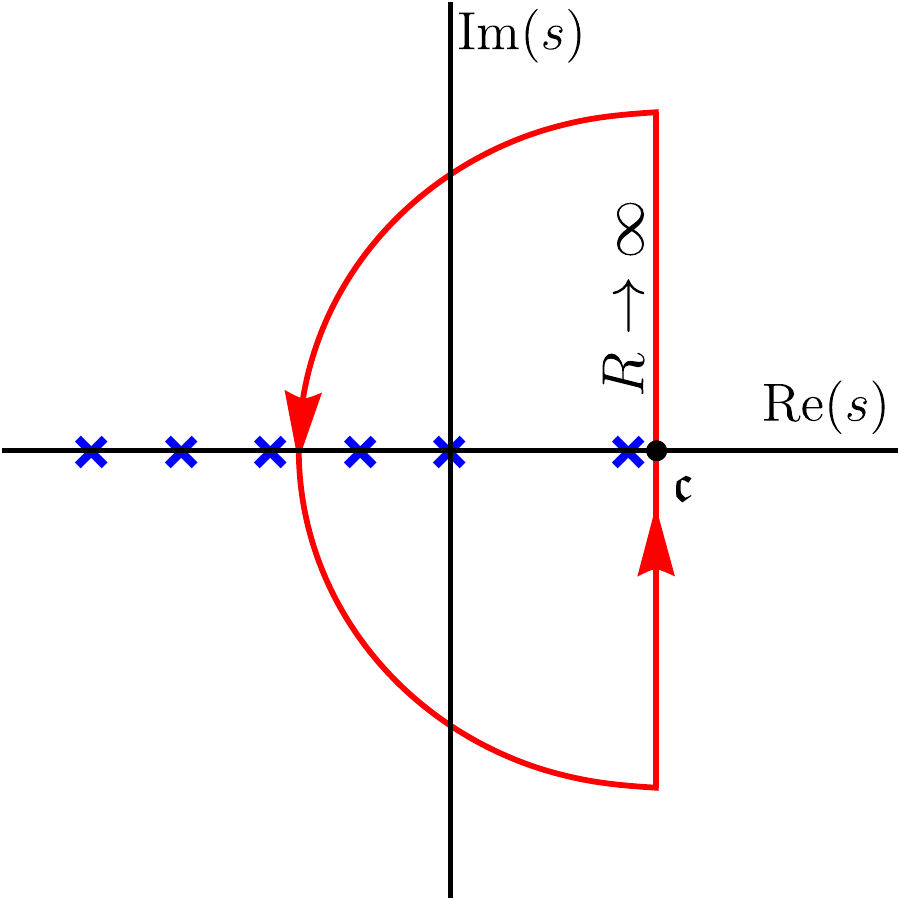}
  \end{center}
  \caption{Poles of the integrand of Eq.~(\ref{intens}). The integration contour $C$ is given by the location of $\mathfrak{c}$. In order to apply the Cauchy's residue theorem, the  semicircle of radius  $R$ is closed on the left.}
  \label{Fig:poles}
\end{figure}

Therefore, given the form of the integrand, the result is convergent if $s/2>1$, which implies $\mathfrak{c}>2$. The latter is implemented by a closed contour with the form depicted in Fig.~\ref{Fig:poles}, which forces to take into account all the poles located in the negative real axis. Thus, the proper application of the Cauchy's residue theorem yields:
\bea
\mathcal{Z}_0&=&\frac{e^{\tilde{\beta}}}{2}-\frac{1}{2}+\frac{1}{2r}\left(\frac{2}{\tilde{\beta}^2}-1\right)\nn\\
&+&\sum_{n=1}^\infty\frac{\left(-\sqrt{2r}\tilde{\beta}\right)^n}{n!}\zeta_H\left(-\frac{n}{2},1+\frac{1}{2r}\right),
\label{Z0final}
\eea
where the ground state $n=0$ was isolated and we define the inverse re-scaled temperature $\tilde{\beta}\equiv mc^2\beta$, and from now, we call its inverse $\tilde{\tau}$ as the system's temperature.

\section{Results}
\begin{figure}
    \centering
    \includegraphics[scale=0.4]{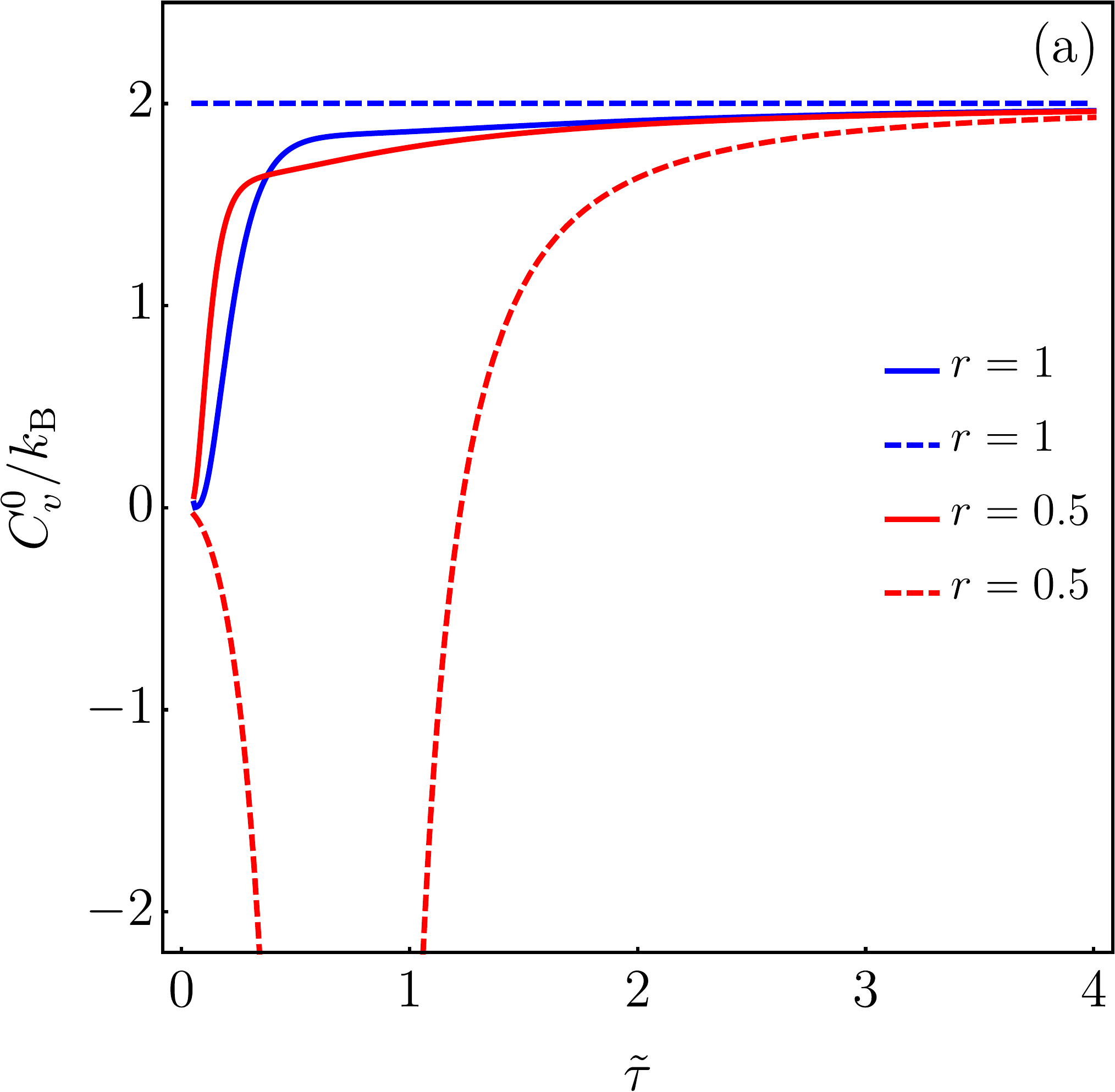}\\
    \vspace{0.5cm}
    \includegraphics[scale=0.4]{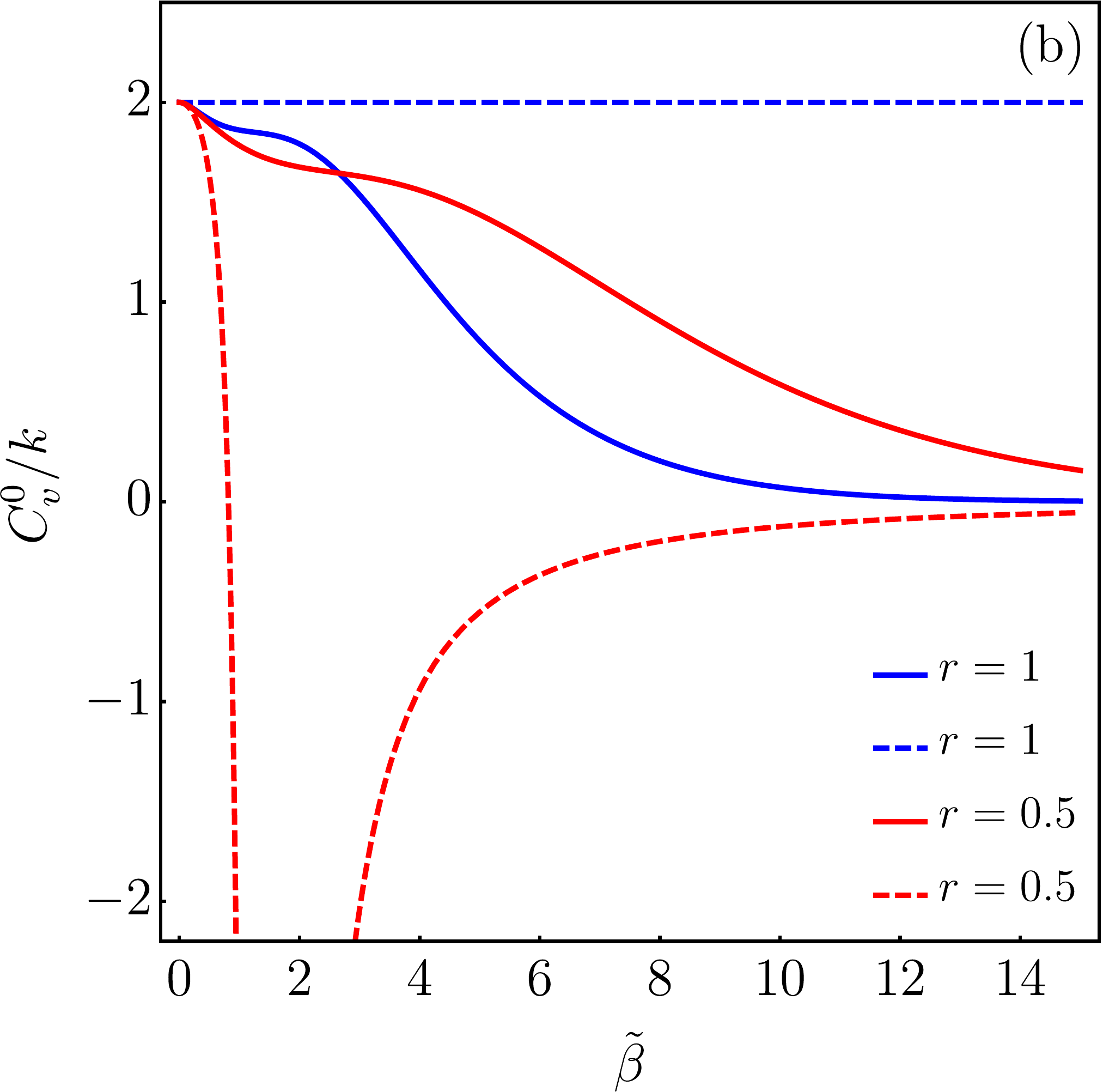}
    \caption{Specific heat $C_v^0$ computed from Eq.~(\ref{Z0final}) (continuous lines) and compared with the results of Eq.~(\ref{Z0B}) (dashed lines) as a function of (a) $\tilde{\tau}$ and (b) $\tilde{\beta}$ for $r=1$ (blue) and $r=0.5$ (red).}
    \label{Fig:Cv0}
\end{figure}

In Boumali's paper, the authors show the functional behavior of the Helmholtz free energy, the average energy, the entropy, and the specific heat.  The last one is of particular interest in this comment, given that for some parameter configurations, their specific heat is not positive definite, and it does not vanish when the temperature goes to zero. To clarify this,  Fig.~\ref{Fig:Cv0} shows the specific heat capacity without the superstatistics prescription $C_v^0$, calculated from the Boumali's  partition function:
\bea
\mathcal{Z}_0^{\text{B}}=\frac{1}{2r\tilde{\beta}^2}+\zeta_H\left(0,\frac{1}{2r}\right),
\label{Z0B}
\eea
and compared with our  results obtained from Eq.~\ref{Z0final}. The results were computed for values of the parameter $r$, which have been studied by several authors, namely, $r=1,\,0.5$. Although the specific heat of the authors has an expected functional shape for some parameter configurations, the construction of $\mathcal{Z}_0$ implies that it has to work for any other parameter selection, nevertheless, as  Fig.~\ref{Fig:Cv0}(a) and (b) show, changing the energy scale (controlled by $r$) implies an $C_v^0$ which does not vanish when $\tilde{\tau}\equiv1/\tilde{\beta}\rightarrow0$, with negative regions, a not well-defined limit for high-temperatures ($\tilde{\beta}\rightarrow 0$) and an apparently singularity. On the other hand, if the specific heat is computed from Eq.~\ref{Z0final} such features are recovered. The latter implies that the results starting from $\mathcal{Z}_0^{\text{B}}$ do not give a proper thermodynamic treatment. 

\begin{figure*}
    \centering
    \includegraphics[scale=0.35]{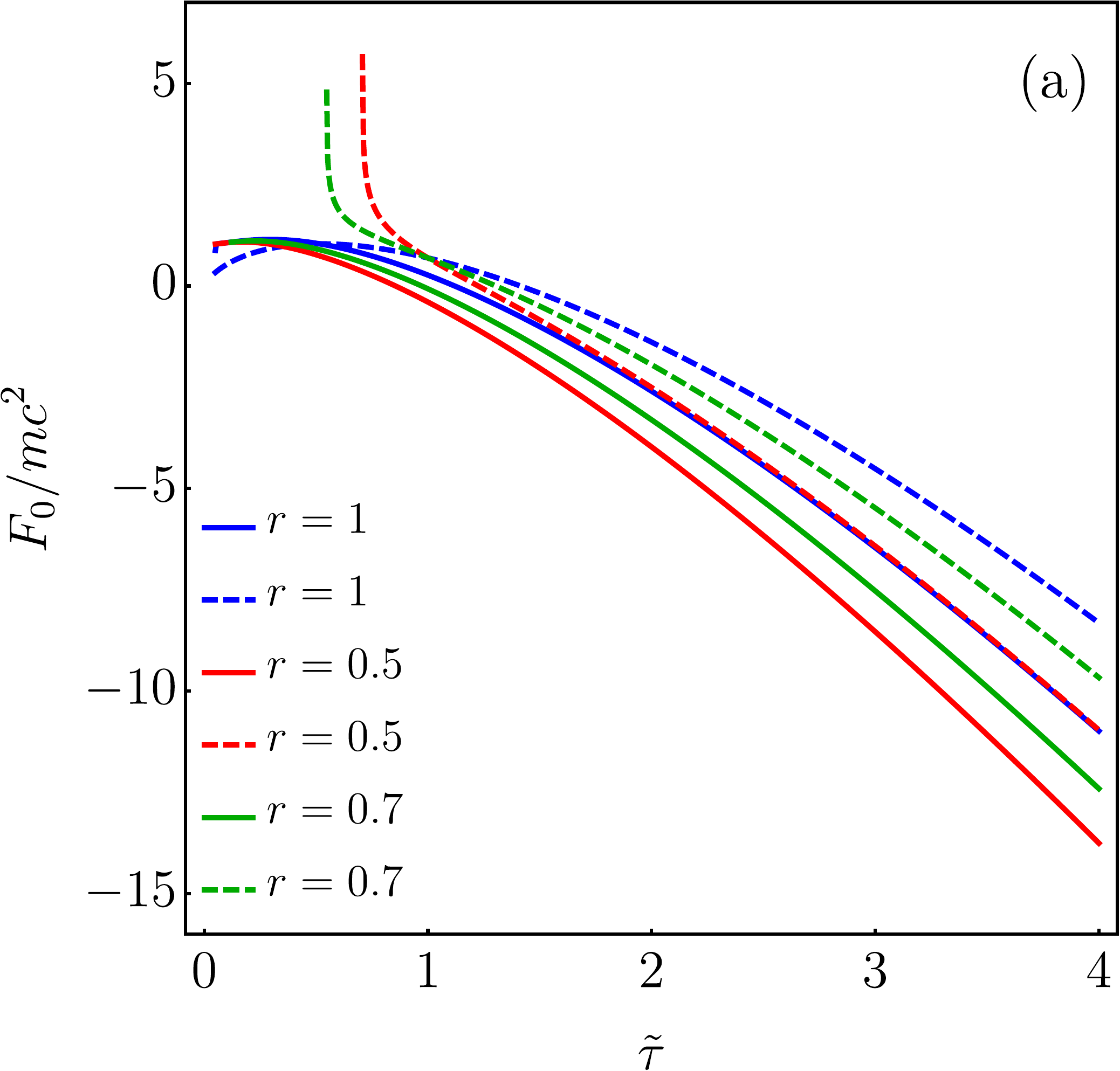}~\quad ~~\includegraphics[scale=0.35]{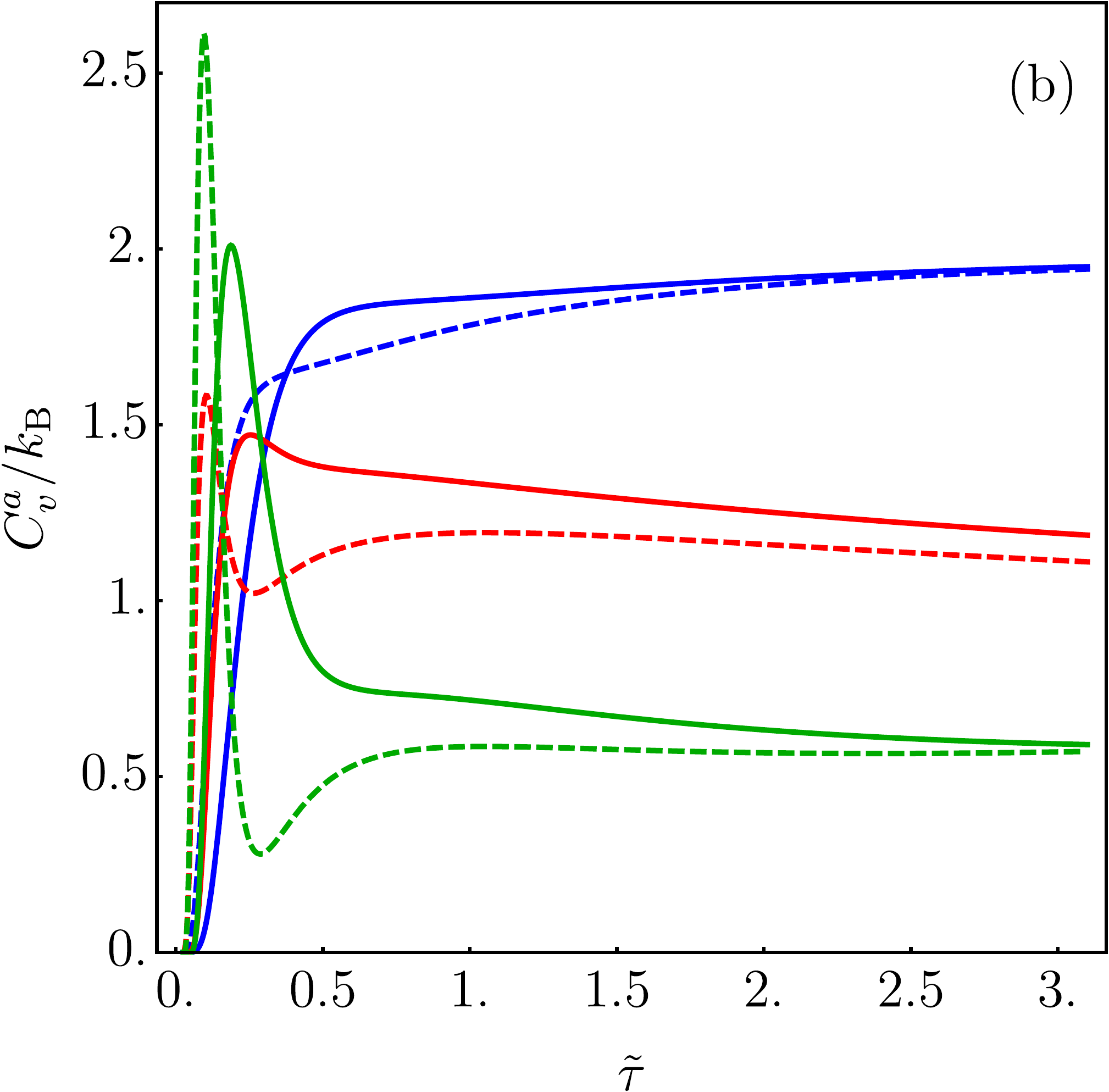}
    \caption{\textbf{(a)}: Helmholtz free energy $F_0$ computed from Eq.~(\ref{Z0final}) (continuous lines) compared with the results obtained from Eq.~(\ref{Z0B}) (dashed lines) for $r=1, 0.5, 0.7$. \textbf{(b)}: Specific heat obtained from the super statistical prescription of Eq.~(\ref{LnqZ_expansion}) for $r=1$ (continuous line) and $r=0.5$ (dashed line). The blue lines correspond to $q=1$, the red to $q=1.1$, and the green to $q=1.2$.}
    \label{Fig:F0}
\end{figure*}

The authors report a series of discontinuities present in $F$, $U$, and $S$, directly impacting the functional form of $C_v$. They argue that undesirable behavior is removed by imposing restrictions over the parameter $q$. To clarify this point, the Helmholtz free energy $F_0$ computed from $\mathcal{Z}_0$ and $\mathcal{Z}_0^{\text{B}}$ is shown in  Fig.~\ref{Fig:F0}(a). Note that the discussed discontinuities are present for values of $r\neq1$, which are related to divergences into the natural logarithm of the partition function. Also, there is not \textit{a priori} argument to establish restrictions over $q$; thus, the logical conclusion is that such non-analytical regions come from a wrong choice of $\mathcal{Z}_0^{\text{B}}$.

Finally, in order to give the result in accordance to Eqs.~(\ref{LnqZ_expansion})-(\ref{CvSSexpanded}), Fig.~\ref{Fig:F0}(b) shows the specific heat capacity as a function of temperature for $r=0.5$,  $r=1$,  and for $q$ values out of the restricted interval that the authors refers. Here, we demonstrate that such values are physically accessible, which is a consequence of the validity of the Tsallis non-extensive statistics for all values of $q$. Also, the shape and analytic behavior of the specific heat validate the choice of $\ln_a\mathcal{Z}$ instead of the common natural logarithm.\\

It is worth to mention that the specific  heat of Fig.~\ref{Fig:F0}(b) shows the so-called Schottky anomaly, i.e.,~a region where the specific heat is not a monotonic function of the temperature. Such anomaly is commonly present in confined systems with external magnetic fields so that if the thermal environment gives to the system an amount of energy close to the transition from the ground state to the first excited state, a significant change in the entropy takes place which is evidenced as a peak in the specific heat.

\begin{table}
    \centering
\begin{tabular}{| c | c | c | c |}
\hline
&$n=0$ & $n=1$ & $\Delta E/mc^2$ \\
\hline
$r=0.5$ & 1   & 1.4142 & 0.4142\\ 
\hline
$r=1$   & 1   & 1.7320 & 0.7320\\
\hline
\end{tabular}
\caption{Energy difference between the ground state and the first excited state for two values of the parameter $r$.}
\label{tablaanomaly}
\end{table}

As an analogy with magnetized systems where the Schottky anomaly has been
extensively
studied~\cite{castano2019super,castano2018comparative,hoi2019schottky,yahyah2019heat,boyacioglu2012heat,PhysRevB.100.174509,shukri2019comprehensive}, note that the energy levels of Eq.~\ref{energylevels} has the form of Landau levels for a fermion in an external constant magnetic field with intensity
$B_0=r$. A change in the magnetic field implies different separation
$\Delta E/mc^2=(E_1-E_0)/mc^2$ of the ground and first excited state. Table~\ref{tablaanomaly}
shows such difference for the values of $r$ presented in Fig~\ref{Fig:F0}(b), and as it can be noticed, the Schottky-like peak is close
around to $\Delta E/mc^2$'s value. The peak is not centered at the temperature
$\tilde{\tau}=\Delta E/(k_Bmc^2)$, because all states contribute to the partition function, not just
the ground and the first excited state. Therefore, deviations from that
temperature value are expected.

\section{Conclusions}

Based on the preceding arguments, we infer that the results for the
superstatistical thermal properties of the one-dimensional Dirac oscillator
presented in Ref.~\cite{boumali2020superstatistical} are mistaken. In the first
instance, the formalism used by the authors did not conserve the Legendre
structure of the thermodynamics related to a Tsallis non-extensive framework,
leading to a wrong description of the observables. The latter is a crucial
point, given that in their original paper, the authors give an analogy with
Tsallis statistics through constrictions over the free parameters in the
$\chi^2$-distribution function, as it was commented in
\ref{Sec:The_superstatistical_formalism_revisited}. Moreover, the canonical
partition function calculation did not consider all the poles in the negative
real axis, which means it was incomplete. Hence, anomalous behaviors in the
thermodynamical properties are observed in the results of the authors.

The results in this work have corrected their calculations by using the $q$-logarithm and implementing a proper partition function for the system. We have demonstrated that anomalies in $F$, $U$, and $S$ come from a wrong choice in partition function, which introduces the divergences observed, instead of an interval of possible values for the parameter $q$ imposed by the choosing of the Gamma distribution. As we have shown, well-behaved thermodynamical functions are obtained out of the interval mentioned by the authors; thus, a restriction of over $q$ is not needed. On the other hand, we show that in the specific heat the Schottky anomaly is present, which can be interpreted as a consequence of an effective magnetic field provided by a Landau-like quantization.

Finally, the authors' analysis of the super statistical properties of graphene is also mistaken, given the fact that it was performed under a wrong formalism and with an incomplete partition function.

\section*{AUTHOR CONTRIBUTIONS}
\textbf{Jorge David Castaño-Yepes}: Conceptualization, Methodology, Formal analysis, Investigation, Data curation, Writing - original draft, Visualization, Supervision, Project administration.\textbf{ I. A. Lujan-Cabrera:} Validation, Formal analysis, Data curation.\textbf{ C.F. Ramirez-Gutierrez: }Conceptualization, Validation, Data curation, Writing - original draft, Visualization.

\bibliography{RevisedManuscript-Proofs}
\end{document}